\documentclass[pra,twocolumn,superscriptaddress]{revtex4}
\usepackage{blindtext}
\usepackage{amsfonts}
\usepackage{amsmath,amssymb}
\usepackage{graphicx}
\usepackage{subfigure}
\usepackage{bm}
\usepackage{epstopdf}
\DeclareMathOperator{\sech}{sech}
\begin{document}
\title{$p-$wave Superfluid Phases of Fermi Molecules in a Bilayer Lattice Array}

\author{G.A. Dom\'{\i}nguez-Castro and R. Paredes} 
\affiliation{Instituto de F\'{\i}sica, Universidad
Nacional Aut\'onoma de M\'exico, Apartado Postal 20-364, M\'exico D.
F. 01000, Mexico.}  

 \email{rosario@fisica.unam.mx}


\begin{abstract}
We investigate the emergence of superfluid ${\bf p}=p_x+ip_y$ phases in an ultracold gas of dipolar Fermi molecules lying in two parallel square lattices in 2D. As shown by a two body study, dipole moments oriented in opposite directions in each layer is the key ingredient in our mean field analysis from which unconventional superfluidity is predicted. The $T=0$ phase diagram summarizes our findings: Stable and metastable superfluid phases appear as a function of the dipole-dipole interaction coupling parameter. A first order phase transition, and thus a mixture of superfluid phases at different densities, is revealed from the coexistence curves in the metastable region. Our model predicts that these superfluid phases can be observed experimentally at $0.6$ nK in molecules of NaK confined in optical lattices of size $a =532$nm.
\end{abstract}

\maketitle

\section{Introduction}
Experimental evidence suggests that Cooper pairing in high $T_c$ superconducting materials, in unconventional superconductors and in superfluid $^3$He, is not in the usual zero angular momentum state, instead, pairs are bound in a $d$-wave, $p-$wave or higher order superconducting order parameter \cite{Leggett75, Tsuei, Volovik, Pfleiderer, Rice, Stewart, Mackenzie, Brison}.  Also from phenomenology, the consensus is that in the particular case of cuprates and ruthenates, the transport with zero viscosity results from the formation of Cooper pairs traveling in planes  \cite{Millis, Leggett}. Such a frictionless transport, characterized by a non $s$-wave energy gap, still remains as an open question because the mechanism that replaces the usual electronic pairing through phonons is unknown. 

As it is also well known, the existent analogies between neutral superfluids and charged superconductors, make ultracold atoms and molecules the best candidates to emulate condensed matter systems behavior. In particular, ultracold polar molecules can be used to access the physics of correlated pairs that form unconventional superfluid phases belonging to a 2D domain. The long-range character of dipolar interaction manifested in producing a contribution of all partial waves at low energies \cite{Levinsen, Baranov, Li, Boudjemaa}, together with its partial attractiveness, open the possibility of producing BCS pairing. These characteristics of dipolar interactions provide the key ingredients to investigate higher superfluid pairing mechanisms through a BCS mean field scheme characterized by the order parameter $\Delta(\mathbf{r}-\mathbf{r'})=V_{dip}(\mathbf{r}-\mathbf{r'}) \langle \hat \Psi(\mathbf{r}) \hat \Psi(\mathbf{r'}) \rangle$. 

The research of unconventional superfluid phases has attracted the interest of the ultracold Fermi gases community for more than a decade. In particular, the search and observation of $p-$wave superfluid phases have become a major research goal. In part, this is motivated because of the potential applications that such systems offer in quantum information processing \cite{Kitaev,Sarma1}. Current experimental facilities allow the design of ad hoc systems in which dc and ac external electric fields can be varied to control both, the strength and anisotropy of dipole-dipole interactions. For instance, very recently the strongly interacting regime in dipolar $^{167}$Er atoms loaded in a 3D optical lattice has been reached \cite{Ferlaino}. As proposed in Refs. \cite{Fedorov1,Fedorov2}, due to the repulsive core between dipoles situated in a bilayer array, the $s-$wave pairing is suppressed and, instead, $p-$wave or higher partial wave superfluid phases arise in molecules with rotational moments $J=0$ and $J=1$.
\begin{figure}[htbp]
\begin{center}
\includegraphics[width=3.3in]{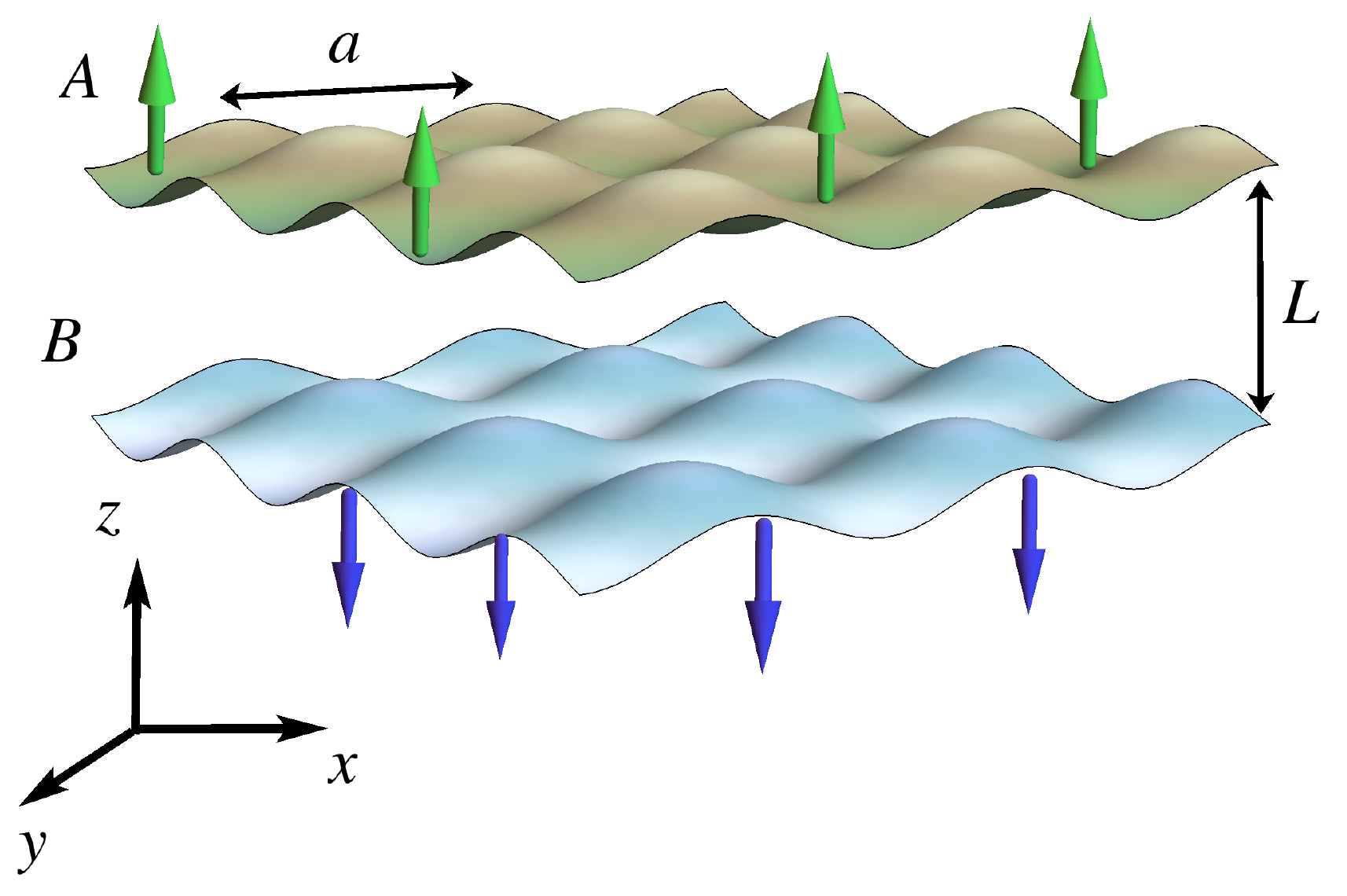} 
\end{center}
\caption{(Color online) Schematic representation of dipolar Fermi molecules situated in a bilayer array composed of parallel optical lattices in two dimensions.}
\label{Fig1}
\end{figure}

In this manuscript we investigate unconventional superfluid phases of ultracold dipolar Fermi molecules and, in particular, we consider a prospect in which antisymmetric pairing may emerge. For this purpose we consider molecules confined in a bilayer array composed of 2D optical lattices having square symmetry, with the molecules oriented in opposite directions, see Fig \ref{Fig1}. 
The strong tail-tail repulsion between molecules lying in on-site positions in layers $A$ and $B$ inhibits pairing formation; in contrast, Cooper pairing is favored when molecules locate at different lattice sites. Thus, superfluidity arises as a consequence of the energy saving associated to a reduction of the electrostatic repulsion. 
It is important to mention that superfluidity of the system here studied is analogous to superconductivity of Sr$_2$RuO$_4$, in the sense that the frictionless transport of pairs has the same origin, namely, electrostatic interactions. Thus, besides the physics of superfluidity that can be addressed, our model can also be used to access the physics of unconventional ruthenate superconductors.   

This paper is organized as follows. In section II we describe the model, then, based on the analysis for scattered pairs and bound states presented in section III, we work in the BCS mean field approach to study unconventional superfluidity in section IV. As we shall demonstrate, the proposed model predicts a first order phase transition between superfluid phases at different densities, as a function of the interaction strength. This is obtained from the stability criteria imposed by the second law of thermodynamics, which allows to recognize stable and metastable phases. The physics of the model is summarized in the phase diagram at zero temperature. In section V we determine the superfluid density tensor with the purpose of establishing the BKT transition temperature. Finally,  we present our conclusions in section VI.

\section{Model}
\label{model}
 We consider polar Fermi molecules placed in two parallel optical lattices in 2D separated by a distance $L$, being the structure of each layer a square lattice of size $a$ lying in $x-y$ plane, $V_{latt}(x,y)=\sin^2 \frac{\pi}{a}x + \sin^2 \frac{\pi}{a}y$. Layers will be labeled as $A$ and $B$. The dipoles are oriented perpendicularly to the layers and in opposite directions in different layers, see Fig. \ref{Fig1}. As delineated in \cite{Fedorov2}, this type of configuration can be created by setting polar molecules with rotational moments $J=0$ and $J=1$ in layers $A$ and $B$ respectively. Then, applying an ac electric field, oriented perpendicularly to the lattices, gives rise to having effective dipole moments pointing in opposite directions \cite{Micheli}. In this scenario the potential of interaction between two molecules situated in different layers is:

\begin{equation}
V_{dip}(r)= -d^2 \frac{(r^2-2L^2)}{(r^2+L^2)^{5/2}},
\end{equation} 
where $-d^2$ is the scalar product of the effective dipole moments in the layers, $L$ and $r$ are the inter-layer and intra-layer separations respectively \cite{Wall}. Molecules can move through sites in a given layer, but they cannot tunnel between layers.
Given the arrangement illustrated in Fig. \ref{Fig1}, several interactions between pairs can arise, being such interactions essential to define the possible phases or ordered structures that can be formed \cite{Ho, Gadsbolle, Potter, Camacho2}. The purpose of this work is to study the $p-$wave superfluidity emerging from attractive interactions between molecules lying in layers $A$ and $B$. Since dipoles are oriented along opposite directions $\hat k$ and $-\hat k$, and due to the discreteness of positions of them in the lattices, the election of interlayer separation is a key parameter to ensure that attractive interactions dominate against attractive or repulsive intra-layer interactions. In this work we shall consider $\Lambda=0.2$, being such dimensionless quantity defined in terms of the lattice spacing $a$ as $\Lambda= L/a$. We notice that for this value of $\Lambda$, the interaction energy between molecules lying in $A$ and $B$ layers is attractive except for $r=0$. 
In addition to the effective dipole-dipole interaction strength $\chi =m_{eff}d^{2}/a\hbar$, with $m_{eff}=\hbar^{2}/2ta^2$ the effective mass, superfluidity will be analyzed in terms of the filling factor $n$.

\section{Two body physics}

Since the robustness of superfluidity lies on the existence of the molecule pair formation, one should warrant on one side that such pairs are not bound, but on the contrary, molecules in each pair belong to states in the continuum, and, on the other side, that pairs contributing to the superposition that defines the many body superfluid state are energetically favorable, that is, that its electrostatic energy is compatible with the condition imposed by minimization of energy.

\begin{figure}[htbp]
\begin{center}
\includegraphics[width=3.0in]{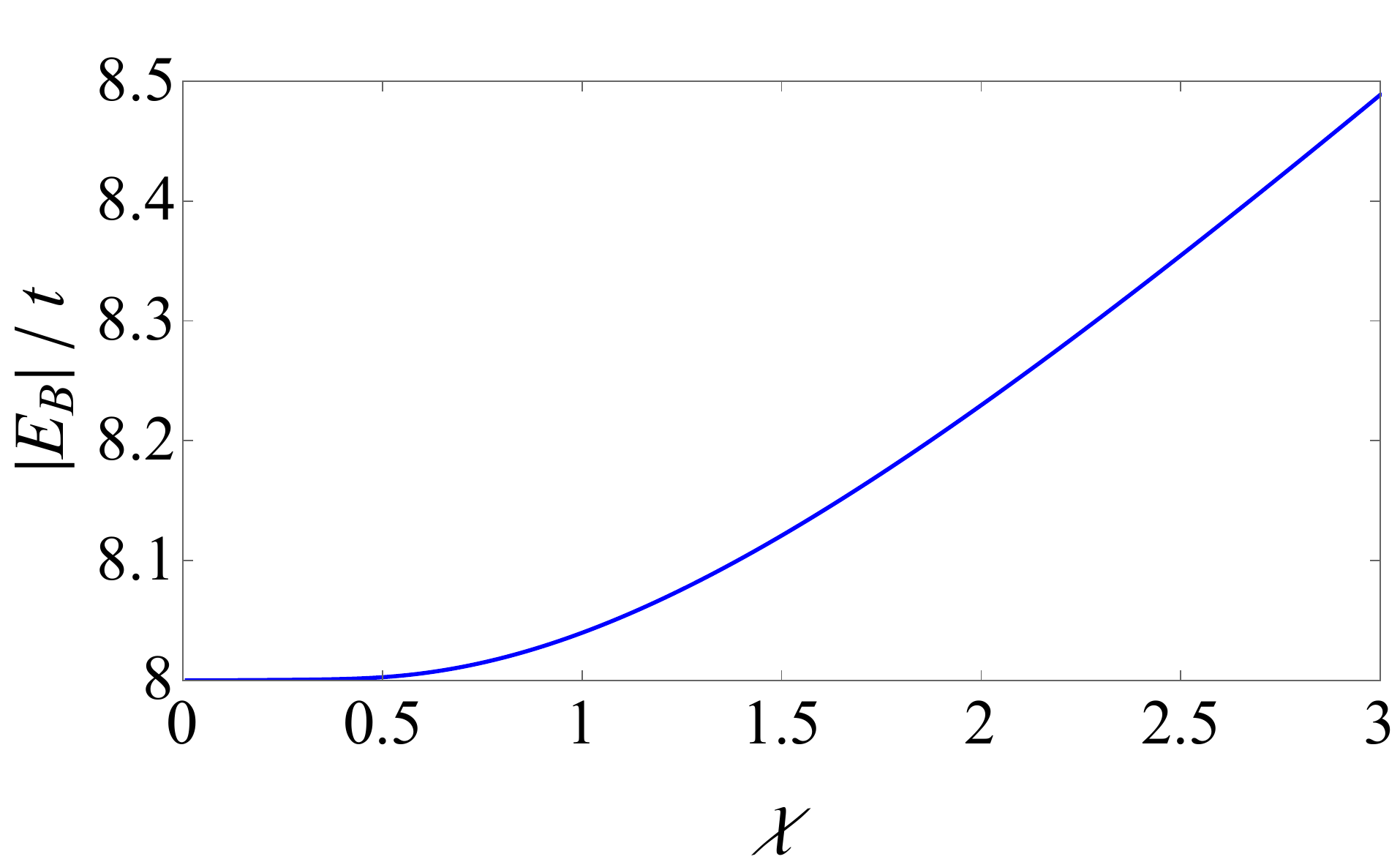} 
\end{center}
\caption{(Color online) Absolute value of the binding energy of a dimer composed of two dipolar Fermi molecules lying in a bilayer array of  squares lattices in 2D. The binding energy $|E_B|$ is plotted as a function of the effective dipole-dipole interaction strength $\chi$. Below $\chi <0.5$ pairs are not bound.}
\label{Fig2}
\end{figure}
To get insight of the appropriate values for which the effective dipole-dipole interaction $\chi$ tends to form either scattered pairs and true bound state pairs, we examine the physics of two molecules. First, we investigate the bound energies $E_B$ as a function of $\chi$. In Fig. \ref{Fig2} we plot the solutions for the binding energy given by the following equation,
\begin{equation}
1= \frac{1}{\Omega}\sum_{\mathbf{q}} \frac{V_{dip}(\mathbf{q})}{E_B-E_{\mathbf{K},\mathbf{q}}}
\end{equation}
being $V_{dip}(\mathbf{q})$ the Fourier transform of the interaction potential, $V_{dip}(\mathbf{q})=\sum_{\mathbf{r}}V_{dip}(\mathbf{r})e^{-i\mathbf{q}\cdot\mathbf{r}}$, with ${\mathbf{q}}$ the relative momentum between two molecules lying in layers $A$ and $B$, and $\Omega$ the number of sites. $E_{\mathbf{K},\mathbf{q}} = -4t(\cos (K_{x}a/2)\cos (q_{x}a) + \cos (K_{y}a/2)\cos (q_{y}a))$, being $\mathbf{K}$ the center of mass vector in the first Brillouin zone. 
As one can see from Fig. \ref{Fig2}, for values of $\chi<0.5$ molecules are not bound, and thus for lower values of $\chi$ we are dealing with purely fermionic physics. Notice that we have plotted $|E_B|/t$ and thus dimeric molecular states exist when $|E_B|$ exceeds $8t$. 

In addition to the examination of the two-molecule binding energy in 2D, a simple analysis in 1D will allow us to delineate several conclusions for the scattered pairs, and in particular to establish which pairs must be considered in the many body picture. For this purpose we solve by exact diagonalization the stationary Schr\"odinger equation,
\begin{equation}\nonumber
H| m \rangle =E_{m}| m \rangle,
\end{equation}
considering just two molecules, with Hamiltonian,
\begin{equation}\nonumber
H=-t\sum_{\langle {i,j} \rangle} (a_{i}^\dagger a_{j} + b_{i}^\dagger b_{j} + h.c.) + \sum_{i,j}V_{dip}(i-j) a_{i}^\dagger b_{j}^\dagger b_{j} a_{i}.
\end{equation}

\noindent
Where $m$ is a simple label to identify the eigenstates, which are linear superpositions of Fock states $| n_1^A,n_2^A, \ldots n_\Omega^A; n_1^B,n_2^B, \ldots n_\Omega^B \rangle$, being $n_i ^{A/B}= 0$ or $1$.
\begin{figure}[htbp]
\begin{center}
\includegraphics[width=3.0in]{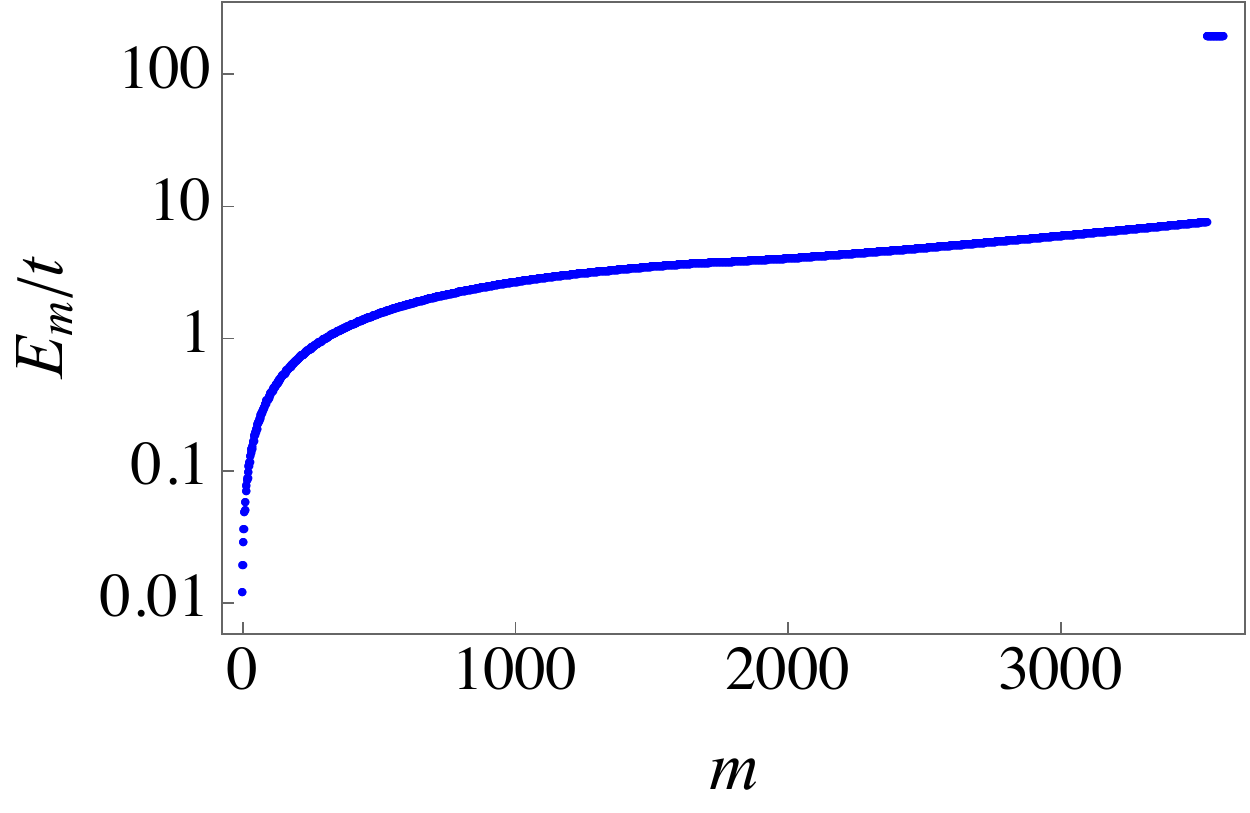} 
\end{center}
\caption{(Color online) Energy spectrum of scattered pairs in 1D for $\chi=0.4$ and $N_x=60$. Each scattered pair is composed of two dipolar Fermi molecules lying in layers $A$ and $B$. Notice that the energy has been shifted by $4t$ to better appreciate the full spectrum in logarithmic scale.}
\label{Fig3}
\end{figure}

\begin{figure}[htbp]
\begin{center}
\includegraphics[width=3.3in]{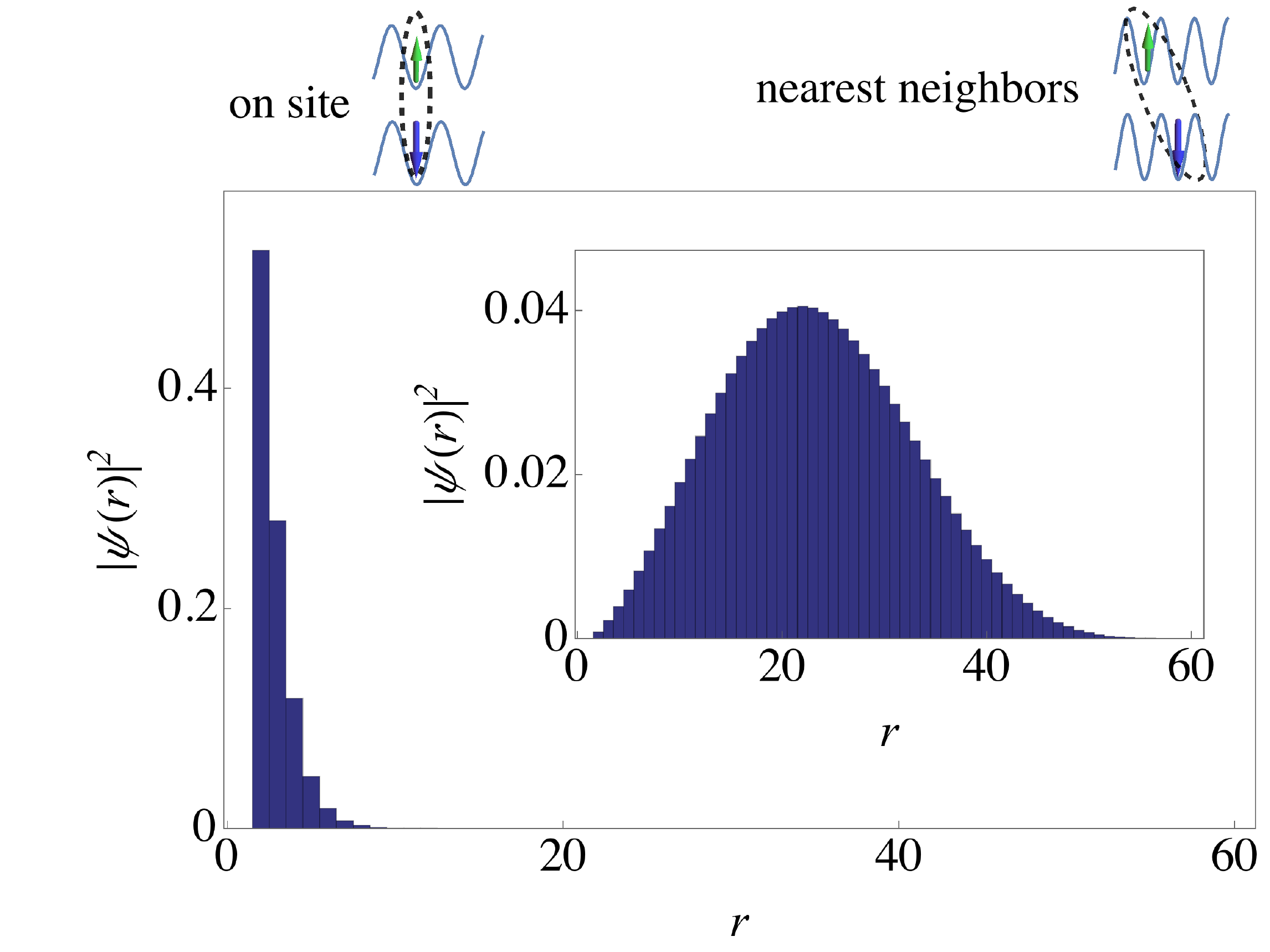} 
\end{center}
\caption{(Color online) Probability distributions of scattered pairs of molecules situated in on-site ($r=0$) and not on-site ($r  \neq 0$) configurations for $\chi=1.8$. The inset represents probability distributions for $\chi=0.4$.}
\label{Fig4}
\end{figure}

In Fig. \ref{Fig3} we plot the energy spectrum for $\chi=0.4$ and $N_x=60$.  To better appreciate the full spectrum the energy has been shifted by $4t$ and plotted in logarithmic scale. As one can see most of the energies lie in the range $[0,8t]$, while a small set of 60 levels have a large energy $E \sim 200t$. The pairs in the first category are associated to scattered pairs in which one of the molecules occupies a given site in layer $A$ and the other, placed in layer $B$, can be find in a nearest neighbor position ($nn$), or next nearest neighbor position ($nnn$), and so on. By contrast, scattered pairs having $E \sim 200t$ correspond to two molecules situated one on top of the other in layers $A$ and $B$, that is, they occupy the same site in both layers (see top of Fig. \ref{Fig4}). We shall call this last configurations as on-site.  We should stress that the energy necessary to form this type of pairs is huge compared to that required to form scattered pairs $nn$, $nnn$ or other configurations. This is why in the many body analysis such pairs in on-site configurations must be neglected. This approximation resembles the one-layer homogeneous case where the hard core repulsive is neglected in the superfluid analysis, but of main importance in suppressing inelastic collisions \cite{Levinsen, Cooper}. As expected, the bottom of the energy spectrum becomes shifted as the value of $\chi$ is increased. In fact, as such effective interaction strength grows, scattered pairs transform into localized dimers, resembling bound states. To illustrate the occurrence of this fact in Fig. \ref{Fig4} we plot the distribution pair configurations lying on-site ($r=0$), $nn$ ($r=1$), $nnn$ ($r=2$), and the successive configurations, for $\chi=1.8$. As can be appreciated from this figure, pair configurations beyond $r=10$ are essentially null. At the inset of the same figure we plot the distribution of pair configurations for $\chi=0.4$, showing that pairs at larger separations exist. This distribution is typical for values of $\chi < 1.4$.

Two main statements can summarize the findings of our two molecule study. The first is that the superfluid state should be investigated considering values of the effective interaction lower than $\chi < 0.5$. The second is that pairs of molecules lying in on-site configurations must be discarded in the $N$ body analysis since the energy cost of these pairs is two orders of magnitude greater than that associated to pairs in other configurations. One can also justify ignoring on-site pairs as a condition compatible with the minimization energy requirement of the BCS scheme. 

\section{Bardeen-Cooper-Schrieffer superfluidity}
Our starting point is the many body Hamiltonian that represents the model depicted in section \ref{model} and considers the  prescriptions, established in the previous section, on dipole-dipole interactions placed in on-site positions.
With this in mind, the effective Hamiltonian of this dipolar Fermi system is given by
\begin{eqnarray}
\label{Heff-i}
H&=&-t\sum_{\langle {\mathbf{i}},{\mathbf{j}} \rangle} (a_{\mathbf{i}}^\dagger a_{\mathbf{j}} + b_{\mathbf{i}}^\dagger b_{\mathbf{j}} + h.c.)\\ \nonumber
&&+ \sum_{{\mathbf{i}} \neq {\mathbf{j}}}V_{dip}({\mathbf{i}}-{\mathbf{j}}) a_{\mathbf{i}}^\dagger b_{\mathbf{j}}^\dagger b_{\mathbf{j}} a_{\mathbf{i}}- \mu \sum_{\mathbf{i}} (a_{\mathbf{i}}^\dagger a_{\mathbf{i}}+ b_{\mathbf{i}}^\dagger b_{\mathbf{i}}),
\end{eqnarray}
here, the nearest-neighbor tunneling strength has been identified with the usual label $t$, operators $a_{\mathbf{i}}$ ($a_{\mathbf{i}}^{\dagger}$) and $b_{\mathbf{i}}$ ($b_{\mathbf{i}}^{\dagger}$) destroy (create) Fermi molecules at sites $\mathbf{i}$ in layers $A$ and $B$ respectively and $\mu$ is the chemical potential. It is important to notice that the effective Hamiltonian (\ref{Heff-i}) has the particle-hole symmetry $(a_{\mathbf{i}}, b_{\mathbf{i}}) \rightarrow (a_{\mathbf{i}}^{\dagger},b_{\mathbf{i}}^{\dagger})$ and $(a_{\mathbf{i}}^{\dagger},b_{\mathbf{i}}^{\dagger}) \rightarrow (a_{\mathbf{i}}, b_{\mathbf{i}})$, where the filling factor transform as $n \rightarrow 1-n$ , the nearest neighbor tunneling $t \rightarrow -t$, and  the chemical potential as $\mu \rightarrow -\mu + \sum_{\mathbf{i}}V_{dip}(\mathbf{i})$.  As described above, due to long range attractive interactions, molecules in different layers can form pairs at low temperatures, particularly $p-$wave pairing is the dominant symmetry as a result of Fermi statistics. To study this superfluid pairing state at $T=0$ we use the self-consistent Hartree-BCS approximation. Due to no interlayer hopping, the Fock contribution to the decoupling of the interaction term is absent. Assuming pairing between $\mathbf{k}$ and $\mathbf{-k}$ states, the mean-field Hamiltonian in momentum representation becomes

\begin{equation}
H = \sum_{\mathbf{k}}\xi_{\mathbf{k}}(a_{\mathbf{k}}^{\dagger}a_{\mathbf{k}} + b_{\mathbf{k}}^{\dagger}b_{\mathbf{k}}) + \Delta_{\mathbf{k}}^{*} b_{- \mathbf{k}}a_{\mathbf{k}} + \Delta_{\mathbf{k}} a_{\mathbf{k}}^{\dagger}b_{-\mathbf{k}}^{\dagger},
\end{equation}
where $\xi_{\mathbf{k}}=\epsilon_{\mathbf{k}}-\mu+nV_{dip}(\mathbf{k}=0)$, $\epsilon_{\mathbf{k}}=-2t(\cos k_{x}a + \cos k_{y}a)$ is the band energy, and $\Delta_{\mathbf{k}}$ is the superfluid order parameter given by $$\Delta_{\mathbf{k}}=\frac{1}{\Omega} \sum_{{\mathbf{k'}}} V_{dip}(\mathbf{k}-\mathbf{k'}) \langle b_{-\mathbf{k'}} a_{\mathbf{k'}}\rangle.$$ 
As standard, a Bogoliubov transformation leads us to obtain coupled equations for the energy gap and the number occupation,

\begin{equation}\label{Delta}
\Delta_{\mathbf{k}}=-\frac{1}{\Omega} \sum_{\mathbf{k'}} V_{dip}(\mathbf{k}-\mathbf{k'})\frac{\Delta_{\mathbf{k'}}}{2E_{\mathbf{k'}}} \tanh \left( \frac{\beta E_\mathbf{k'}}{2} \right),
\end{equation}

\begin{equation}\label{ff}
n_A=\frac{1}{2} \left[ 1-\frac{1}{\Omega} \sum_{\mathbf{k}} \frac{\xi_{\mathbf{k}}}{E_{\mathbf{k}}} \tanh \left( \frac{\beta E_\mathbf{k}}{2} \right) \right],
\end{equation}
where the quasiparticle energy spectrum is $E_{\mathbf{k}}= \sqrt{\xi_{\mathbf{k}}^2 + |\Delta_{\mathbf{k}} |^2}$ and $\beta= 1/(k_B T)$, with $k_B$ the Boltzmann constant. To calculate the energy gap as a function of the effective dipole-dipole interaction $\chi$ and the filling factor $n$, we consider the physics provided by the two body analysis of the previous section. That is, the $N$-body problem will be studied for $\chi \in[0,0.5]$ and $n \in [0,1]$. 

In our calculations we solve self-consistently Eqs. \eqref{Delta} and \eqref{ff}, considering that each lattice has $\Omega=121 \times 121$ sites, and equal filling factors $n_A=n_B$ in each layer.

\subsection{$p_x+ ip_y$ superfluid pairing}
\begin{figure}[htbp]
\begin{center}
\includegraphics[width=3.2in]{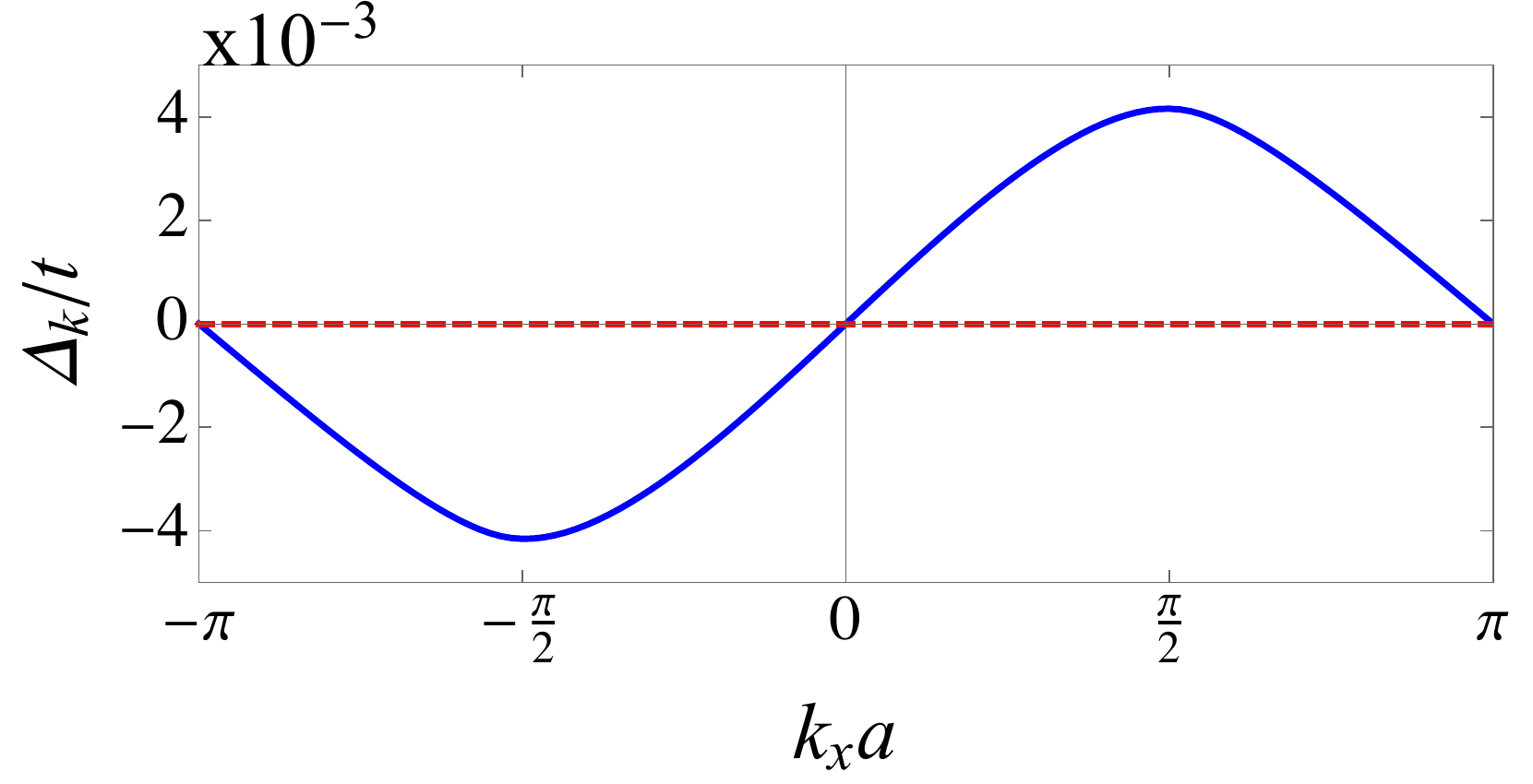} 
\includegraphics[width=3.2in]{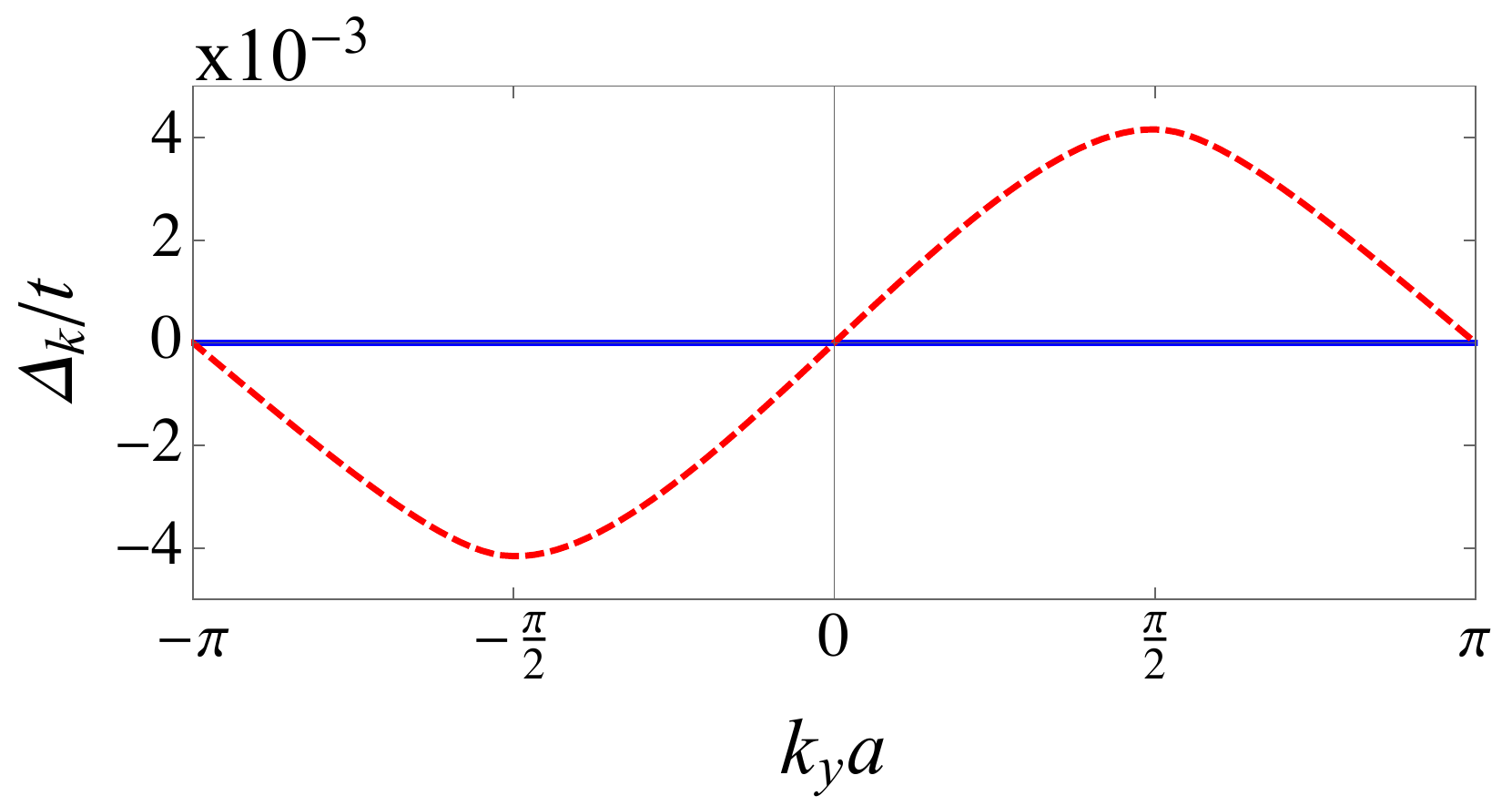} 
\end{center}
\caption{(Color online) Energy gap parameter $\Delta_\mathbf{k}$ as a function of $\mathbf{k}$ for filling factor $n=0.16$, coupling strength $\chi=0.4$ and $T=0$.  Upper and lower panels correspond to $k_y a=0$ and $k_x a=0$ respectively.}
\label{Fig5}
\end{figure}

At $T=0$ Eqs.  \eqref{Delta} and \eqref{ff} leads to a gap energy $\Delta_\mathbf{k}$ that exhibits an antisymmetric behavior characteristic of $p_x+ ip_y$ pairing. For illustration purposes in Fig. \ref{Fig5} we show the obtained energy gap parameter for a coupling strength $\chi=0.4$, and a filling factor $n=0.16$. Solid and dashed lines correspond respectively to real and imaginary parts of $\Delta_\mathbf{k}$, plotted as a function of $k_x a$ and $k_y a$ In the upper figure we plot $\Delta_\mathbf{k}$ considering $k_y a=0$, conversely, lower figure corresponds to $k_x a=0$. As can be appreciated from these figures our model supports an almost pure $l=1$ superfluid phase since $\Delta_\mathbf{k}  \propto \sin k_x a + i \sin k_y a$.  
As found by \cite{Anderson} the solution $p_x +ip_y$ is the candidate for the most stable $p$-wave pairing, in contrast to $p_z$ symmetry \cite{Book}. Even more, the $p_x +ip_y$ pairing breaks time-reversal symmetry and it is a class $D$ topological superfluid with Majorana modes at its edges.

\begin{figure}[htbp]
\begin{center}
\includegraphics[width=3.2in]{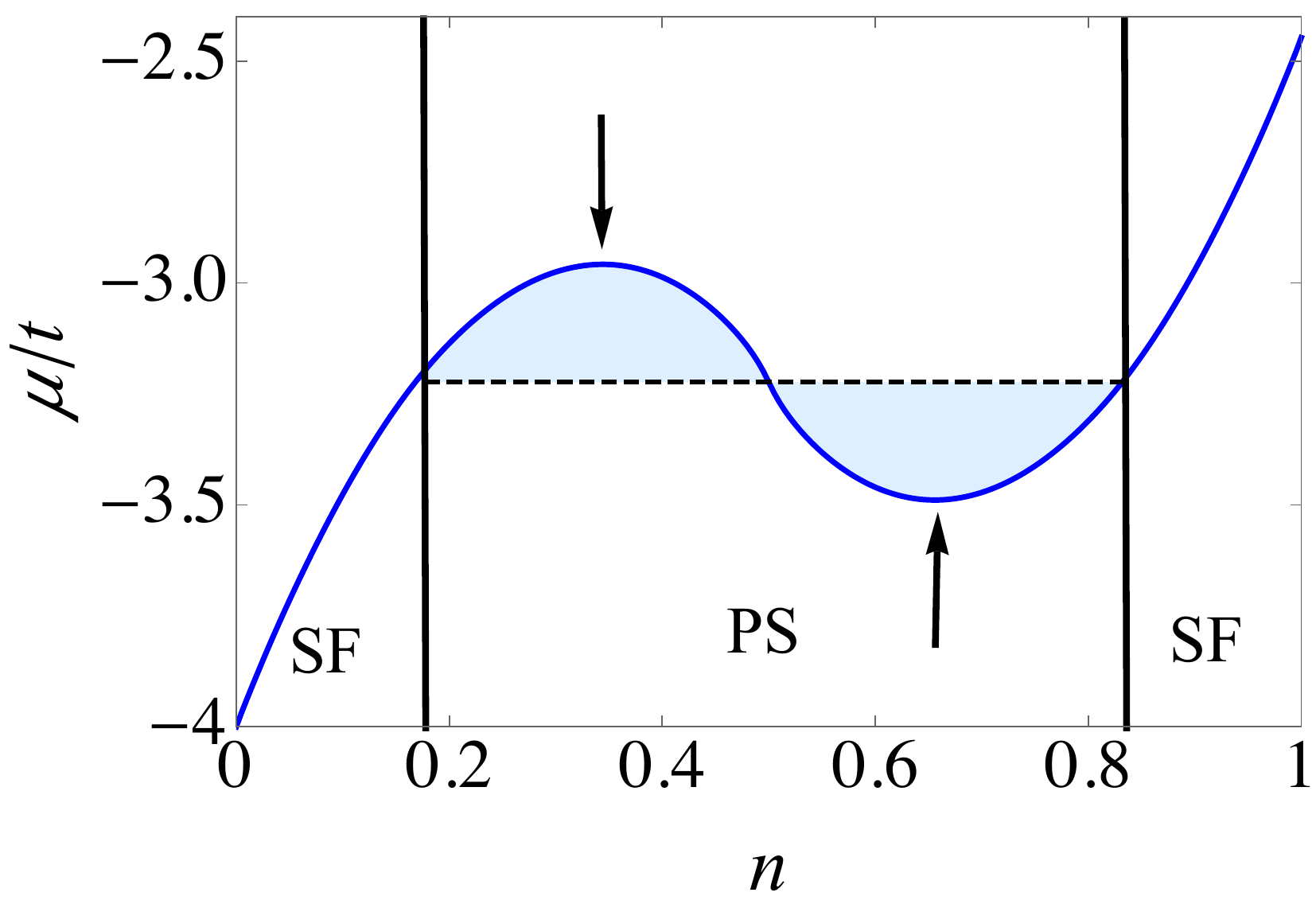} 
\end{center}
\caption{(Color online) Chemical potential $\mu$ vs filling factor $n$ at $T=0$, and interaction strength $\chi=0.4$. Black solid line separates stable from metastable $p$-wave superfluid phases. For values of filling fractions inside the range indicated with black arrows }
\label{Fig6}
\end{figure}

When the interaction strength or the filling factor are large enough, the system becomes unstable towards phase separation. In order to study this behavior in Fig. \ref{Fig6} we plot the chemical potential as a function of the filling factor $n$ for interaction strength $\chi=0.4$ at zero temperature. These results can be compared with those made for single layer models \cite{Liu,Midtgaard} where Pauli exclusion principle prevents particles to occupy the same lattice site. Several conclusions can be made from the information encoded in Fig. \ref{Fig6}. Stability criteria for the compressibility $\kappa=1/n^2 (\partial n/\partial \mu)$ demands $\kappa>0$, thus, the collection of points having negative derivative in the curve $\mu$ vs. $n$ cannot exists. Arrows at maximum and minimum values of $\mu$ indicate the range of $n$ for which the system is unstable. The phase separation region can be recognized from the Maxwell construction of equal areas, however due to the particle-hole symmetry, such construction simplifies to determine the filling factor $n_1$, which satisfies the following condition, $\mu(n_1)=\mu(n=0.5)=\mu(1-n_1)$. Blue shaded region  in Fig. \ref{Fig6} illustrates Maxwell construction of equal areas. Solid black lines separate stable and metastable SF phases, that is, within the region labeled PS -phase separation- the system becomes a mixture of SF phases at different densities.

\subsection{Phase diagrams}

\begin{figure}[htbp]
\begin{center}
\includegraphics[width=3.1in]{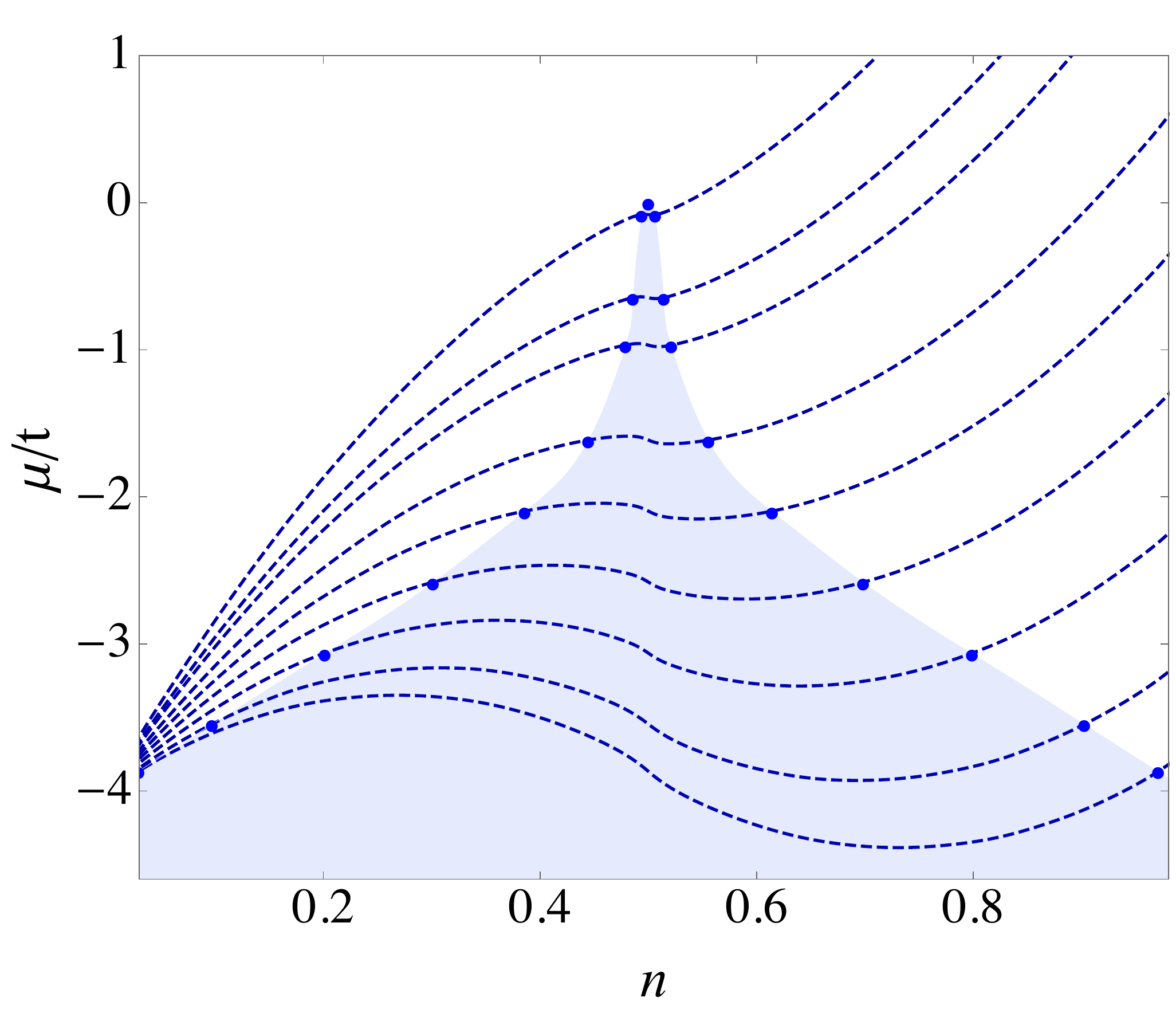} 
\end{center}
\caption{(Color online) Dotted curves correspond to chemical potential $\mu$ as a function of the filling factor $n$ at $T=0$, for different values of interaction strength $\chi$, the values of this parameter from bottom to top are 0.48, 0.44, 0.38, 0.32, 0.26, 0.2, 0.12, 0.07 and 0.01 respectively. Blue shaded region bounds metastable and unstable states. }
\label{Fig7}
\end{figure}

In Fig. \ref{Fig7} we illustrate several ``iso-interaction" curves with the purpose of establishing the coexistence curve of our model. That is, values of $\mu$ for which the state is composed of superfluids at different densities. The blue shaded region bounds the metastable and unstable states referred before. The iso-interaction curves allow us to distinguish the first order transition. The behavior of the derivative of the chemical potential with respect to filling factor suggest that a second order phase transition must occur at half filling and $\chi=0$, however, as we show below, the phase separation region shrinks at half filling and $\chi \rightarrow 0$, that is phase separation no longer exists in this case. 

\begin{figure}[htbp]
\begin{center}
\includegraphics[width=3.4in]{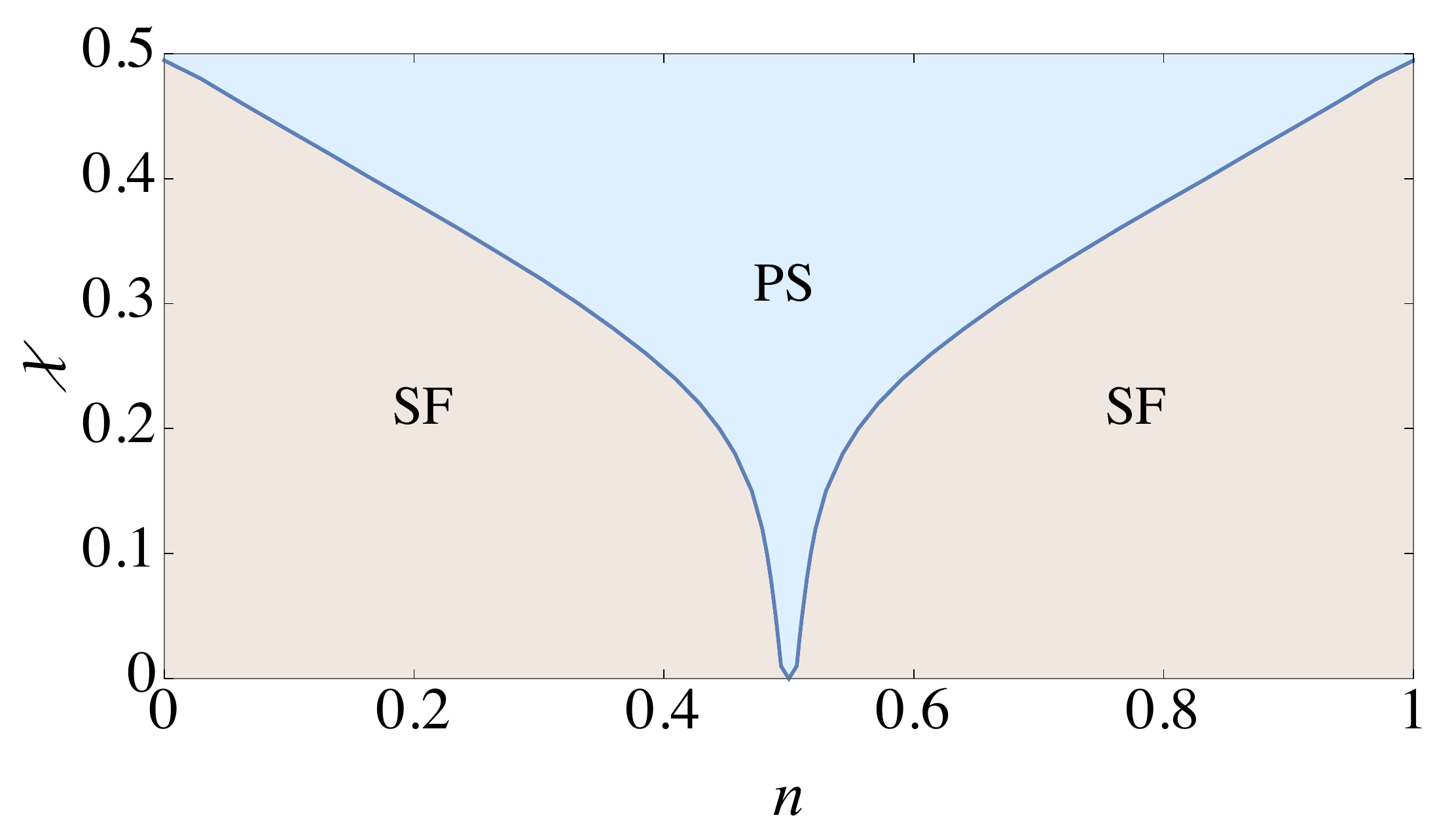} 
\end{center}
\caption{(Color online) Dimensionless interaction coupling parameter $\chi$ vs filling factor $n$ at $T=0$. Brown shaded regions identify stable superfluid phases, while blue surface indicate the values of $\chi$ and $n$ for which the system becomes phase separated and thus a mixture of superfluids at two densities coexist.}
\label{Fig8}
\end{figure}

In Fig. \ref{Fig8} we show the phase diagram of our model as a function of $\chi$ and $n$. There one can observe both, the regions where stable superfluid phases exist, as well as regions where the system is unstable and consequently the superfluid become fragmented in superfluid phases having different densities.
At a critical interaction of $\chi \approx 0.494$ the phase separation occurs between $n=0$ and $n=1$. When the interaction is above this critical value the system becomes a mixture of $n=0$ and $n=1$ densities, and no superfluid phases occur since the stable states are the zero particle or zero hole densities which are the well known insulating states.

\section{Berezinskii-Kosterlitz-Thouless superfluidity}

As it is well known, in two dimensional systems no long range order of conventional type emerges, instead, a BKT finite temperature transition characterized by topological ordering \cite{Berezinskii,Kosterlitz1,Kosterlitz2}, thus the $p-$wave superfluid phase predicted by our model must be of this type. To determine the critical temperature at which the dipolar Fermi system becomes a BKT superfluid we proceed as delineated in \cite{ACamacho}. By introducing a gauge transformation on the system Hamiltonian written in the space representation; $(a_{\mathbf{i}}^{\dagger}, a_{\mathbf{i}})$,  $(b_{\mathbf{i}}^{\dagger}, b_{\mathbf{i}})$ $\rightarrow$ $(a_{\mathbf{i}}^{\dagger}e^{i \mathbf{\theta} \cdot \mathbf{r_i}}, a_{\mathbf{i}}e^{-i \mathbf{\theta} \cdot \mathbf{r_i}})$,  $(b_{\mathbf{i}}^{\dagger}e^{i \mathbf{\theta} \cdot \mathbf{r_i}}, b_{\mathbf{i}}e^{-i \mathbf{\theta} \cdot \mathbf{r_i}})$, being $\mathbf{\theta}=\left( \frac{\theta_x}{aN_x}, \frac{\theta_y}{a N_y}  \right)$, it is possible to recognize that the density separates in both, normal and superfluid components. The superfluid density can be determined as the difference between the normal and superfluid free energies,
\begin{equation}
\rho_{\alpha,\alpha'}=  \lim_{\Theta \rightarrow 0} \frac{1}{Nt}\frac{ F_{\Theta} -F_0}{ \Theta_{\alpha} \Theta_{\alpha'} }=\frac{1}{Nt}\frac{\partial^2 F_{\Theta}}{\partial \Theta_{\alpha} \partial \Theta_{\alpha'}}, \, \alpha=\{x,y\},
\end{equation}
then, after performing a series expansion up to second order in $\theta_x$ and $\theta_y$ one can find the final expression for the superfluid density tensor,
\begin{equation}
\rho_{\alpha\alpha'}=\frac{1}{a^2\Omega}\sum_{\mathbf{k}}\left(n(\mathbf{k})\frac{\partial^2\epsilon_{\mathbf{k}}}{\partial k_{\alpha}\partial k_{\alpha'}}-Y(\mathbf{k})\frac{\partial \epsilon_{\mathbf{k}}}{\partial k_\alpha}\frac{\partial \epsilon_{\mathbf{k}}}{\partial k_{\alpha'}}\right), \label{sfden}
\end{equation}
being $\alpha, \alpha' =x,y$, $n(\mathbf{k})$ the momentum distribution and $Y(\mathbf{k})=\beta \sech ^2(\beta E_{\mathbf{k}}/2)/4$ the Yoshida function. 
\begin{figure}[htbp]
\begin{center}
\includegraphics[width=3.2in]{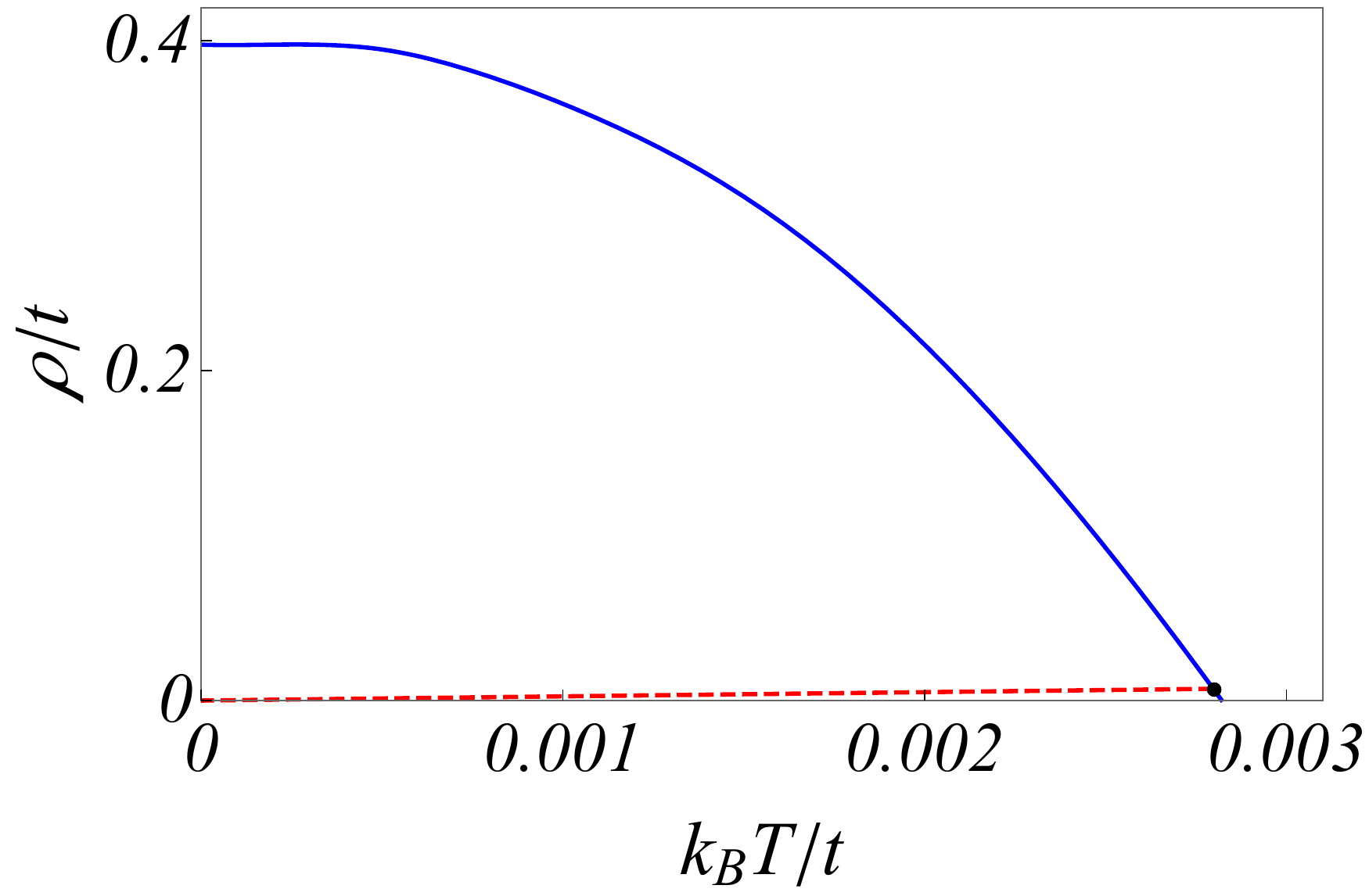} 
\end{center}
\caption{(Color online) Blue solid line is the superfluid density as a function of temperature obtained from BCS mean field scheme. The intersection of the red dashed line and blue curve corresponds to the critical temperature at which BKT transition occurs for an effective  dipole-dipole interaction strength $\chi=0.4$ and a filling factor $n=0.16$.}
\label{Fig9}
\end{figure}
Since the off-diagonal matrix elements of the superfluid density tensor are negligible, the superfluid density is approximately given by $\rho \approx (\rho_{xx}+\rho_{yy})/2$. In order to estimate the critical temperature one has to impose the following BKT condition \cite{Berezinskii,Kosterlitz1,Kosterlitz2},
\begin{equation}
\rho(T_{BKT})= \frac{8}{\pi} k_B T_{BKT} .
 \label{BKT}
 \end{equation}

In Fig. \ref{Fig9} the blue solid line represents the superfluid density as a function of temperature, while red dashed line represents the right hand side of Eq. \ref{BKT}. We find that the crossing between blue curve and red dashed line takes place at $k_{B}T_{BKT}=0.0028t$. For dipolar Fermi molecules of NaK \cite{Park} confined in an optical lattice having a wave length $\lambda \approx 1064$ and an lattice depth of $V_{0}= 5E_r$ \cite{Bloch} the estimated critical transition temperature is $T_{BKT} \approx 0.6$ nK. Despite the fact this temperature is very low, it is important to stress that this bilayer system is intended to be realized by using subwavelength lattices \cite{Yi, Nascimbene, Gullans}, where we may have $L \approx 50$ nm. In this scenario, the critical temperature for a Fermi gas of NaLi molecules, under the same condition $V_{0}= 5E_r$ is about $T_{BKT} \approx 4$ nK.

\section{Conclusions}
We have investigated $p-$wave superfluidity in a model of dipolar Fermi molecules confined in two parallel optical lattices in 2D, separated at a fixed distance. The dipole moments of molecules oriented perpendicular to the layers in opposite directions in different layers give rise to repulsive interactions for molecules situated at the same lattice site, while experiencing an attractive interaction in other cases. The nature of such an electrostatic interaction, together with the attributes of dipolar interactions, are responsible of ${\bf p}=p_x+ip_y$ superfluid phases in our model. To determine the existence of superfluid phases we first addressed the two body physics and demonstrated that, depending on the value of the effective interaction coupling between molecules, two body bound states or scattered states are formed. The many body physics analyzed within the BCS mean field scheme leads us to conclude that resulting from both, long range and anisotropic character of dipolar interactions, the energy gap parameter shows an antisymmetric behavior, characteristic of ${\bf p}=p_x+ip_y$ phases. The phase diagram at $T=0$ shows stable and metastable phases, being the latter a mixture of superfluid phases at different densities, thus providing evidence of a first order phase transition between both superfluids.  We also estimate the BKT transition temperature of the ${\bf p}=p_x+ip_y$ superfluid state, considering recent experimental advances in dipolar ultracold Fermi molecules of NaK, being $T_{BKT}=0.6$nK. Compared with other $p-$wave proposals, the model here discussed represents a genuine candidate to address the physics of the ruthenates, since it contains the strong on-site repulsion, which is an essential ingredient in unconventional superconductivity of those compounds. In addition, both, our model and ruthenate compounds, share the peculiarity of being composed of layered structures. 

\acknowledgements{We acknowledge useful discussions with A. Camacho-Guardian and V. Romero-Rochin. GADC acknowledges CONACYT scholarship. This work was partially funded by grant IN105217 DGAPA (UNAM) and 255573 CONACYT.
}

\end{document}